\documentclass{aa}  
\usepackage{graphicx}
\usepackage{txfonts}
\usepackage{tabularx}
\usepackage{soul}
\usepackage[colorlinks=true, citecolor=blue]{hyperref}
\usepackage{natbib,twoopt}
\bibpunct{(}{)}{;}{a}{}{,}             %% natbib format for A&A and ApJ
\usepackage{hyperref}
\hypersetup{
    colorlinks=true,
    linkcolor=blue,
    citecolor = blue,
    filecolor=magenta,
    urlcolor=blue}
\usepackage{booktabs}

\begin{document}

\authorrunning{Lanzoni et al.}
\titlerunning{Observed nCRD parameters for GC internal dynamics}

\title{New parameters for star cluster dynamics: observational results}

\author{B. Lanzoni\inst{1}\fnmsep\inst{2}, F. R. Ferraro\inst{1}\fnmsep\inst{2}, E. Vesperini\inst{3}}

   \institute{Dipartimento di Fisica e Astronomia, Universit\`a degli Studi di Bologna, Via Gobetti 93/2, I-40129 Bologna, Italy  \email{barbara.lanzoni3@unibo.it}
   \and
   INAF, Osservatorio di Astrofisica e Scienza dello Spazio di Bologna, Via Gobetti 93/3, I-40129 Bologna, Italy
         \and
        Dept. of Astronomy, Indiana University, Bloomington, IN 47401, USA\\
         }

\abstract
{We recently used a large set of Monte Carlo simulations of globular clusters (GCs) to define new fully empirical parameters (named $A_5$, $P_5$, and $S_{2.5}$) able to trace the internal dynamical evolution of dense stellar systems. These parameters are specifically designed to quantify the steepness of the cumulative radial distribution of stars in the innermost region of the host system, which tends to progressively increase with dynamical aging due to core contraction. Following the original definitions, here we measure $A_5$ and $P_5$ in a sample of 40 Galactic GCs homogeneously surveyed through HST photometric observations. In agreement with the predictions of our simulations, the largest values of $A_5$ and $P_5$ are found for the most dynamically evolved GCs, i.e., those previously classified as post-core collapse systems based on the shape of their density profile, and those characterized by the shortest central relaxation times. Moreover, the new dynamical parameters here measured strongly correlate with $A^+_{rh}$, another fully empirical, independent parameter that traces the dynamical age of star clusters through the level of central segregation of blue straggler stars.}

\keywords{Star clusters; Globular star clusters; Dynamical evolution}

\maketitle

\section{Introduction} 
\label{intro}
In recent years special attention has been devoted to characterize the internal dynamical evolution of star clusters. In this respect, globular clusters (GCs) are particularly interesting because they are collisional stellar systems, where two-body interactions among their stars progressively bring the cluster toward a thermodynamically relaxed state in a timescale (the relaxation time) that can be significantly shorter than the cluster age (see, e.g., \citealt{meylan+97}).  Because of such interactions, heavy stars tend to progressively sink toward the central region of the cluster (due to dynamical friction), while low-mass stars migrate outwards and are preferentially lost from the system. This yields a progressive contraction of the central regions (in particular, of
the core radius, $r_c$) and a corresponding increase of the central density ($\rho_0$), until a core-collapse (CC) phase that is eventually halted by the energy provided by primordial or dynamically formed binary stars (see, e.g., \citealp{goodman89, mcmillan+90, gao+91, vesperini+94, heggie+06, trenti+07}). The subsequent post-core-collapse (PCC) phase can be characterized by core oscillations, with several stages during which the cluster rebounds toward a structure with lower density and more extended core, followed by episodes of core contraction and central density increase \citep[see, e.g.,][]{heggie+09}.

The characteristic timescale of such dynamical evolution depends on the cluster internal structure and on the interplay between internal dynamical processes and the effects of the host galaxy tidal field \citep[see, e.g.,][]{heggie+03}.
This implies that Galactic GCs can be in very different stages of their internal dynamical evolution (i.e., they can have very different ``dynamical ages''), in spite of being all coeval, with formation epochs dating back to $\sim 12-13$ Gyr ago \citep[e.g.,][]{ying+25}. The full characterization of any GC therefore requires the knowledge not only of its stellar population content, its overall structure and kinematics, but also of its dynamical age.

\begin{figure*}[h!]
    \centering
  \includegraphics[width=1\linewidth]{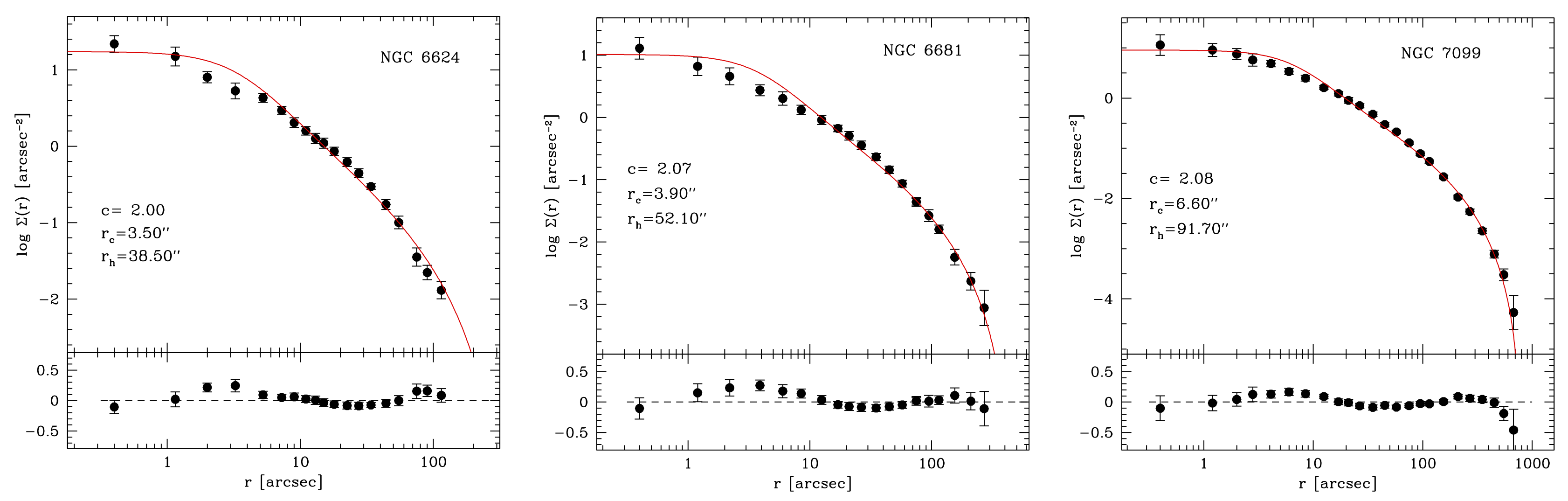}
    \caption{Projected density profile corrected for Galactic field contamination (black circles) for the PCC clusters NGC 6624, NGC 6681, and NGC 7099 (left, central and right panels, respectively). In each case, the values of the best-fit \citet{king66} model parameters are labeled, and the bottom panel shows the residuals between the model and the observations.}
\label{profPCC}
\end{figure*} 

Unfortunately, it is still very hard to understand from observational properties how far a GC is from CC, or  whether it already experienced it and currently is in the gravothermal oscillation phase. The traditional way to pinpoint PCC clusters is to search for a steep central cusp in the surface brightness or projected density distribution, in contrast with the centrally-flat profile commonly observed in non-PCC systems and properly described by \citet{king66} models. However, this method suffers from both operational drawbacks (for instance, an insufficient angular resolution may prevent the detection of the cusp), and from interpretative uncertainties (for instance, the central cusp is expected to significantly reduce during PCC oscillations). In the last decades increasing attention has been therefore devoted to search for parameters able to trace and possibly measure the internal dynamical evolution of star clusters (see, e.g., \citealt{baum+03, bianchini+16, bianchini+18, tiongco+16, webb+17}, for methods requiring the determination of the stellar-mass function or internal kinematics at different radial distances from the center). In this context, we followed an approach based on quantities that are relatively easy to measure from observational data. For instance, we used the radial distribution of blue straggler stars (BSSs) within the host cluster. This is a  special class of stars that, in the color-magnitude diagram (CMD), appear to be systematically brighter than the cluster main-sequence turn-off (MS-TO), thus being easily distinguishable from the other populations. They are thought to be formed through stellar collision, stellar mergers, or mass transfer in binary systems (see, e.g, \citealt{mccrea64, hills_day76, sills+05, perets+09}), which make them more massive than normal cluster stars, with values of $\sim 1.2 M_\odot$ in Galactic GCs (e.g., \citealp{shara+97, gilliland+98, fiorentino+14, raso+19}; see also \citealp{geller_mathieu2012} for results in open clusters). Indeed, signatures of the BSS anomalous formation processes have been detected in their observational properties, such as peculiar rotational velocities (e.g., \citealt{lovisi+10, lovisi+12, ferraro+23a, mathieu_geller09, billi+23, billi+24}), anomalies in their surface chemical abundances ({\citealt{ferraro+06, lovisi+13}), UV-excess in their spectral energy distributions (e.g., \citealt{gosnell+14, gosnell+15, reggiani+25}), and atypical specific frequencies (\citealt{knigge+09, ferraro+26}). Because of their relatively high mass and their ubiquity, BSSs can be used as effective gravitational test particles: their radial distribution and their level of segregation toward the cluster center have been found to be powerful indicators of internal dynamical evolution \citep{ferraro+12, ferraro+20}, with the exciting possibility that their photometric properties also keep memory of the CC event \citep{ferraro+09, dalessandro+13, beccari+19, portegies19, cadelano+22}. 
In particular, the level of BSS central segregation has been  quantified through the $A^+_{rh}$ parameter, defined as the area between the cumulative radial distribution of BSSs and that of a lighter reference population, both measured out to the half-mass radius $r_h$ \citep{alessandrini+16, lanzoni+16}. As expected for an efficient indicator of internal dynamics, the $A^+_{rh}$ parameter shows a strong positive correlation with the number of central relaxation times suffered by the host cluster since formation, both in the Galaxy \citep{ferraro+18, ferraro+23b}, and in the Magellanic Clouds \citep{ferraro+19, dresbach+22, giusti+24, giusti+25}. More recently, we explored the possibility of using the entire stellar population, in place of BSSs only, to trace the host cluster dynamical evolution. By using a large set of MOCCA \citep[e.g.,][]{giersz+13} Monte Carlo simulations of GCs run from different initial conditions, we found \citep{bhat+22, bhat+23, bhat+24} that the level of internal dynamical evolution of the host system is quantified by three parameters named $A_5$, $P_5$, and $S_{2.5}$ tracing the shape of the normalized cumulative radial distribution (nCRD) of cluster stars (see Section \ref{para} below). The present paper is devoted to measure $A_5$ and $P_5$ in a sample of 40 Galactic GCs surveyed with HST photometry, to verify the effectiveness of this new method for the determination of the level of internal dynamical evolution from observational data. 

\section{Definition of the nCRD parameters}  
\label{para}
The nCRD parameters $A_5$, $P_5$, and $S_{2.5}$ have been originally defined in \citet{bhat+22} and further discussed in \citet{bhat+23, bhat+24}. They are specifically designed to quantify the morphology of the nCRD traced by the cluster stellar population in the inner region of the host system. All the details can be found in those papers, but we here briefly summarize the main characteristics.

The population used to build the nCRDs in \citet{bhat+22, bhat+23, bhat+24} is made of all the stars brighter than $V_{\rm cut}=V_{\rm MS-TO} + 0.5$ (with $V_{\rm MS-TO}$ being the magnitude of the MS-TO in the $V$ band), and located within a projected distance $R_{\rm cut}= 0.5\times r_h$ from the cluster center.
This turned out to be the best compromise between the need of maximizing the morphological differences among the nCRDs of clusters that reached different stages of internal dynamical evolution, and providing large enough statistics. By construction, in each cluster the nCRD is a monotonically increasing curve ranging between 0 in the center, and 1 at $R/r_h = 0.5$. The steeper it is, the higher is the concentration of the surveyed stars in innermost regions. To quantify the nCRD steepness we defined the three following parameters (see also Figure 6 in \citealp{bhat+22} for a graphical representation): 
\begin{itemize}
    \item $A_5$ is the area subtended by the nCRD within the innermost 5\% of the half-mass radius (i.e., out to $R/r_h = 0.05$);
    \item $P_5$ is the value of the nCRD (i.e., the fraction of stars) at the same distance from the center;
    \item $S_{2.5}$ is the slope of the tangent to the nCRD at $R/r_h = 0.025$, determined as the straight-line tangent to the third-order polynomial function best-fitting the nCRD. 
\end{itemize} 
By definition, the higher is the concentration of stars within $0.05\times r_h$ (for $A_5$ and $P_5$) or within $0.025\times r_h$ (for $S_{2.5}$), the steeper is the nCRD and the larger is the value of the corresponding nCRD parameter. Since internal dynamical evolution leads to a progressive contraction of the core and therefore tends to increase the central density, steeper nCRDs are expected in the innermost regions of CC and PCC clusters, with respect to less dynamically evolved systems. Indeed, from the analysis of almost 100 simulated GCs, we found increasingly large values of the nCRD parameters for increasingly more dynamically evolved systems, thus demonstrating that these are powerful tracers of internal dynamical evolution \citep[see][hereafter B24]{bhat+24}. As detailed in that paper, the adopted Monte Carlo simulations have been run by assuming different initial masses (number of particles), concentrations, galactocentric distances, binary fractions, black hole retention fractions, and tidally filling factors (see Table 1 in B24), and all GCs have been analyzed after 13 Gyr of evolution. The values of their King concentration parameters, core radii, half-mass radii, and central relaxation times are in very good agreement with those observed in the Milky Way (see Fig. 3 in B24), demonstrating that they are well representative of the Galactic GC population.

\begin{figure*}[!h]
    \centering
  \includegraphics[width=1\linewidth]{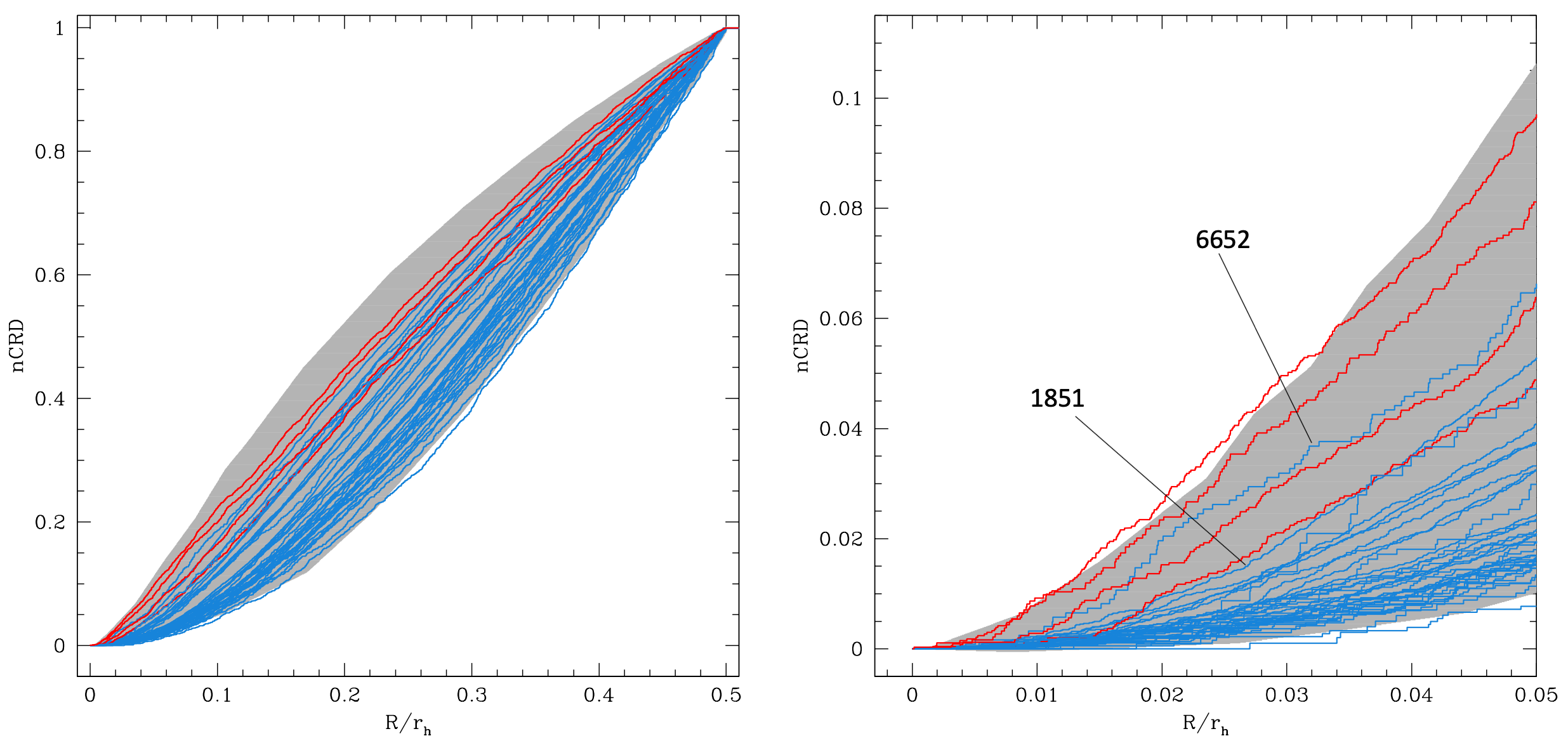}
    \caption{Observed nCRDs of the 40 GCs analyzed in this work. The red lines refer to the four PCC systems in the sample, the blue lines to the other GCs. The right panel shows a zoom in the innermost region (out to $0.05 \times r_h$), where the nCRD parameters are measured. In both panels, the gray shaded areas correspond to the region occupied by the nCRDs of the B24 simulated clusters.}  
\label{ncrd}
\end{figure*}

\section{Measuring the nCRD parameters in observational data}  
\label{data}
Because of the definitions above, the measure of the nCRD parameters in real GCs requires high-resolution photometric observations with an adequate level of completeness, to guarantee the appropriate exploration of the central regions in even high-density systems. For this reason, we took advantage of the data secured in the Hubble Space Telescope UV Legacy Survey of Galactic GCs, providing WFC3/UVIS photometry in the F275W and F336W filters for more than 50 Milky Way GCs \citep[see][]{piotto+15}. As discussed in many previous papers (e.g., \citealt{ferraro+97, ferraro+01, ferraro+03, raso+17}), UV-band observations are particularly effective for the exploration of the innermost regions of high-density, old star clusters, because red giant stars, which dominate the emission in the optical bands, are faint at these wavelengths. This sensibly reduces the problems due to photometric blends and crowding, resulting in completeness levels above 70-80\% in the MS-TO region even for the most crowded clusters, as demonstrated by the artificial star experiments discussed in \citet{ferraro+18}. 

Such a large sample of data all acquired in the same photometric bands also offers the opportunity to  perform a homogeneous selection of the cluster stellar populations needed to build the nCRDs. As discussed above (Section \ref{para}), in the analysis of the simulated clusters we considered all the stars brighter than 0.5 magnitudes below the MS-TO level in the $V$ band. This choice was driven by the fact that reaching fainter optical magnitudes in the innermost regions of high-density clusters is typically very hard, even with HST data (see, e.g., \citealp{raso+19}). However, \citet{bhat+22} also discussed the effect of adopting a fainter cut ($V_{\rm cut} = V_{\rm MS-TO} + 2$) to increase the statistics of the considered sample. They found that, in spite of a slightly reduced sensitivity (due to inclusion of stars with smaller masses, which are less affected by the dynamical processes occurring in the cluster center), the overall trend of the nCRD parameters with dynamical evolution remains unchanged (see their Fig. 9). 
In the present study we use UV magnitudes (in place of optical ones) and we select all the stars that are brighter than 1 (in place of 0.5) magnitude below the MS-TO in the F275W filter, thus achieving larger statistics.
From the analysis of BaSTI isochrones \citep{pietrinferni+13}, we found that adopting a cut at 1 magnitude below the MS-TO results in the inclusions of stars that are just 0.02 M$_\odot$ less massive than those corresponding to a 0.5 magnitude cut, independently of the cluster metallicity. Hence, also considering the findings of \citet{bhat+22} discussed above, we expect that the different choices adopted here have a negligible effect on the results, keeping high the capacity of the nCRD parameters to efficiently trace dynamical evolution.

\begin{figure}
\includegraphics[width=\linewidth]{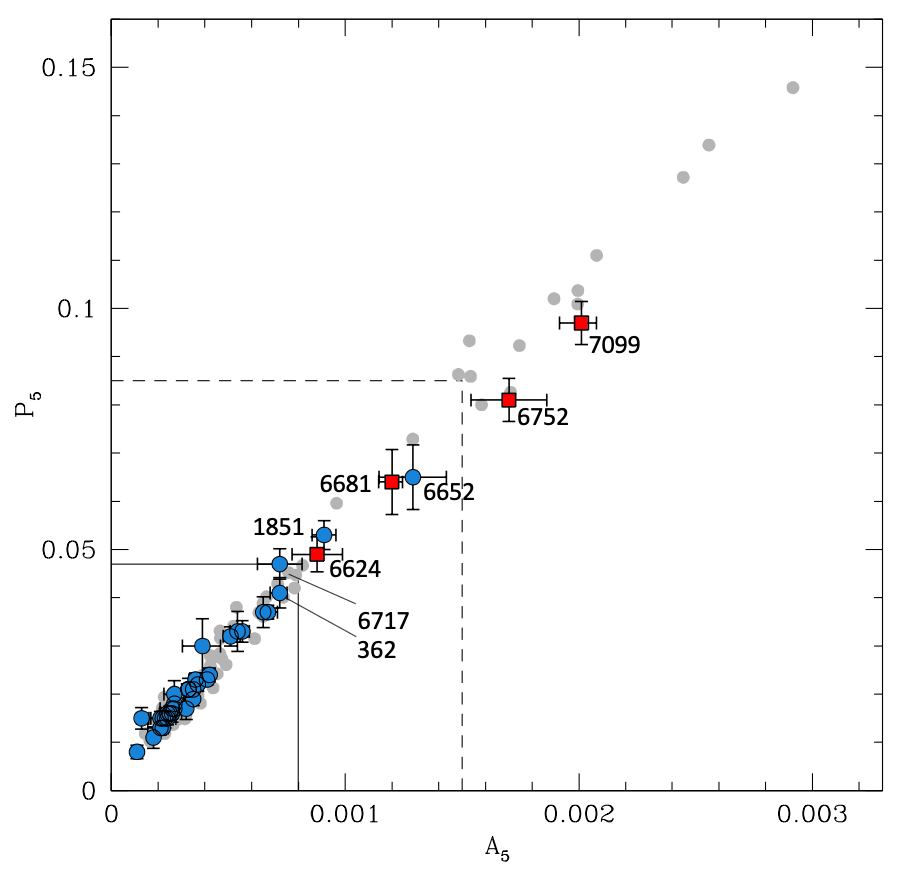}
    \caption{$P_5$ parameter plotted as a function of $A_5$ for the 40 surveyed GCs, with the blue circles referring to pre-CC systems, and the red squares marking PCC clusters. The gray circles show the results from the Monte Carlo simulations of B24. The dashed lines mark the limits in $A_5$ and $P_5$ proposed by B24 as boundaries to separate pre-CC from PCC clusters. The solid lines mark the revised limits proposed here.}
\label{a5p5}
\end{figure}
To guarantee homogeneity among the surveyed clusters, avoiding the combination of datasets acquired in different photometric bands and with different instruments, we excluded from the analysis all the systems where the region included within $R_{\rm cut}=0.5\times r_h$ is not entirely sampled by the WFC3/UVIS field of view ($\sim 162\arcsec\times 162\arcsec$). The adopted values of $r_h$ have been taken from the \citet{harris96} compilation or determined in previous works from the \citet{king66} models that best-fit the observed density profiles (see \citealt{miocchi+13, lanzoni+16, ferraro+18, ferraro+26}). The only exceptions are NGC 6624, NGC 6681, and NGC 7099 (M30), three PCC clusters for which the King concentration has been arbitrarily fixed at $c=2.5$ in the \citet{harris96} catalog, because the presence of a central density cusp prevents the model to properly fit the observations. Here we used the WFC3/UVIS data combined with \citet{stetson+19} catalogs to redetermine the projected density profile of these three clusters from resolved star counts (see, e.g., \citealp{miocchi+13} for the details about the adopted method) and, following the prescriptions of \citet{bhat+22}, we searched for the King model that best reproduces the entire density distribution, without (arbitrarily) excluding the central, cuspy, portion of the profile. The results are shown in Figure \ref{profPCC}. Although the fits are not optimal (as expected and already observed in all PCC clusters), they are reasonably accurate,\footnote{We tried to adopt the same strategy also for the PCC clusters NGC 6397 and NGC 7078 (M15). However, no acceptable King model fits to the observed density profiles could be found, and we therefore excluded these systems from the analyzed sample.} and we therefore adopted as half-mass radii of these systems the values obtained in this way. Table \ref{tab1} lists the values of $r_h$ adopted for the 40 GCs analyzed in this work.

\section{Results}  
\label{resu}
The nCRDs traced by the stars brighter that 1 magnitude below the MS-TO in the F275W filter, and included within a projected distance of $0.5\times r_h$ from the center of the 40 studied clusters are shown as colored thick lines in Figure \ref{ncrd}. The comparison with the region occupied by nCRDs of the simulated clusters (gray shaded area, from B24) shows a very good agreement. 
In addition, we find that the nCRDs of the clusters classified as PCC systems in the \citet{harris96} compilation (red lines) are centrally steeper than the others. This is also in agreement with the theoretical results of B24, and it testifies a higher concentration of stars in the innermost cluster regions due to more advanced dynamical stages. Interestingly, the nCRDs of NGC 1851 and NGC 6652 (see labels) show central steepness comparable to those of NGC 6681 and NGC 6624, which are classified as PCC clusters.  

\begin{figure}
\includegraphics[width=\linewidth]{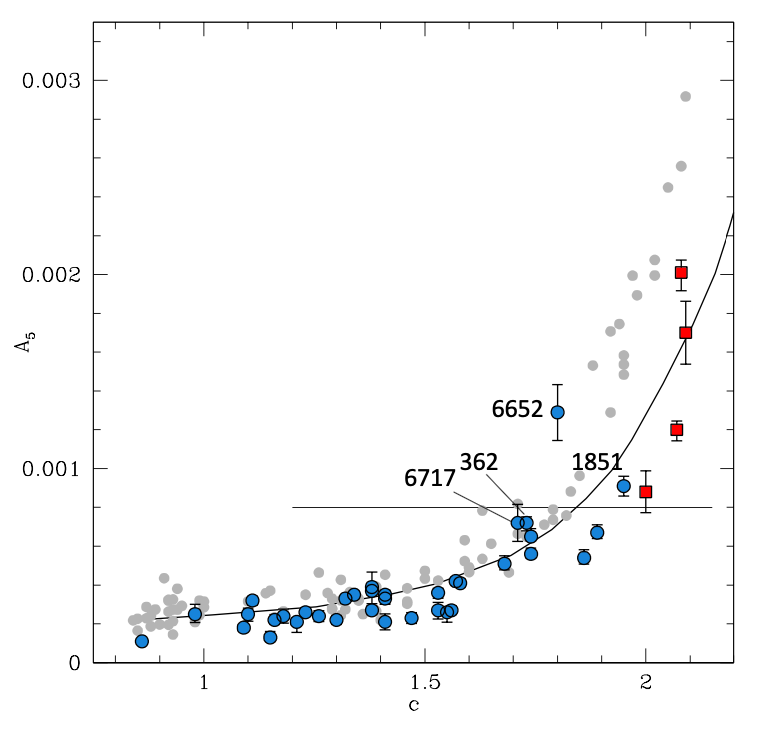}
    \caption{$A_5$ parameter as a function of the best-fit King model concentration $c$ for the observed clusters (colored symbols, as in Fig. \ref{a5p5}) and for the B24 simulations (gray circles). The solid horizontal line marks the proposed $A_5$ boundary separating pre-CC from PCC systems. The thick black curve traces the values of $A_5$ calculated from the integration of King density profiles of varying concentration parameters $c$.}
\label{conc}
\end{figure}

Following the definitions discussed in Sect. \ref{para}, we measured the values of $A_5$ and $P_5$ from each observed nCRD. We did not attempt to also determine the $S_{2.5}$ parameter because several clusters have to few stars observed within $0.025\times r_h$ from the center (for instance, less than 30 stars are counted in half of the GC sample), which prevents a reliable determination of the slope of the nCRD. On the other hand, $S_{2.5}$ shows the same behavior with dynamical evolution as the other two parameters (see B24). We estimated the uncertainty associated to $A_5$ and $P_5$ ($\epsilon_{A5}$ and $\epsilon_{P5}$, respectively) by taking into account the uncertainty on the half-mass radius \citep[see][]{bhat+22} and that on the position of the cluster center. Unfortunately, the error on $r_h$ is not always known 
(e.g., in the Harris catalog) and we therefore adopted an uncertainty of 5\% for all the values of $r_h$, since this is the typical value found in the available cases. Then, we built the nCRDs adopting radial cuts $R_{\rm cut, err} = 0.5\times (r_h \pm5\% r_h)$, and we measured the values of the  $A_5$ and $P_5$ parameters from these nCRDs. The uncertainties are calculated as the absolute differences between the value of $A_5$ obtained by adopting $R_{\rm cut}$ and those obtained by assuming $R_{\rm cut, err}$ for every cluster, and analogously for $P_5$. Miscentering is another potential source of error because the nCRDs change shape if they are built with respect to different positions within the cluster. To estimate this contribution, we adopted the published uncertainties on the center coordinates, if available, or we assumed $0.5\arcsec$ as conservative error. Then, we built the nCRDs with respect to four different positions obtained by offsetting the center to the north, east, south, and west directions by an amount equal to the assumed error. The average absolute difference between the values of $A_5$ (and $P_5$) obtained from each direction and that computed with respect to the right center is finally adopted as the miscentering contribution. The final error on the two parameters ($A_5$ and $P_5$) is computed by summing in quadrature the respective contribution from the uncertainty on $r_h$ and that from miscentering. The obtained values are listed in Table \ref{tab1}. 

\begin{figure*}
    \centering  \includegraphics[width=1\linewidth]{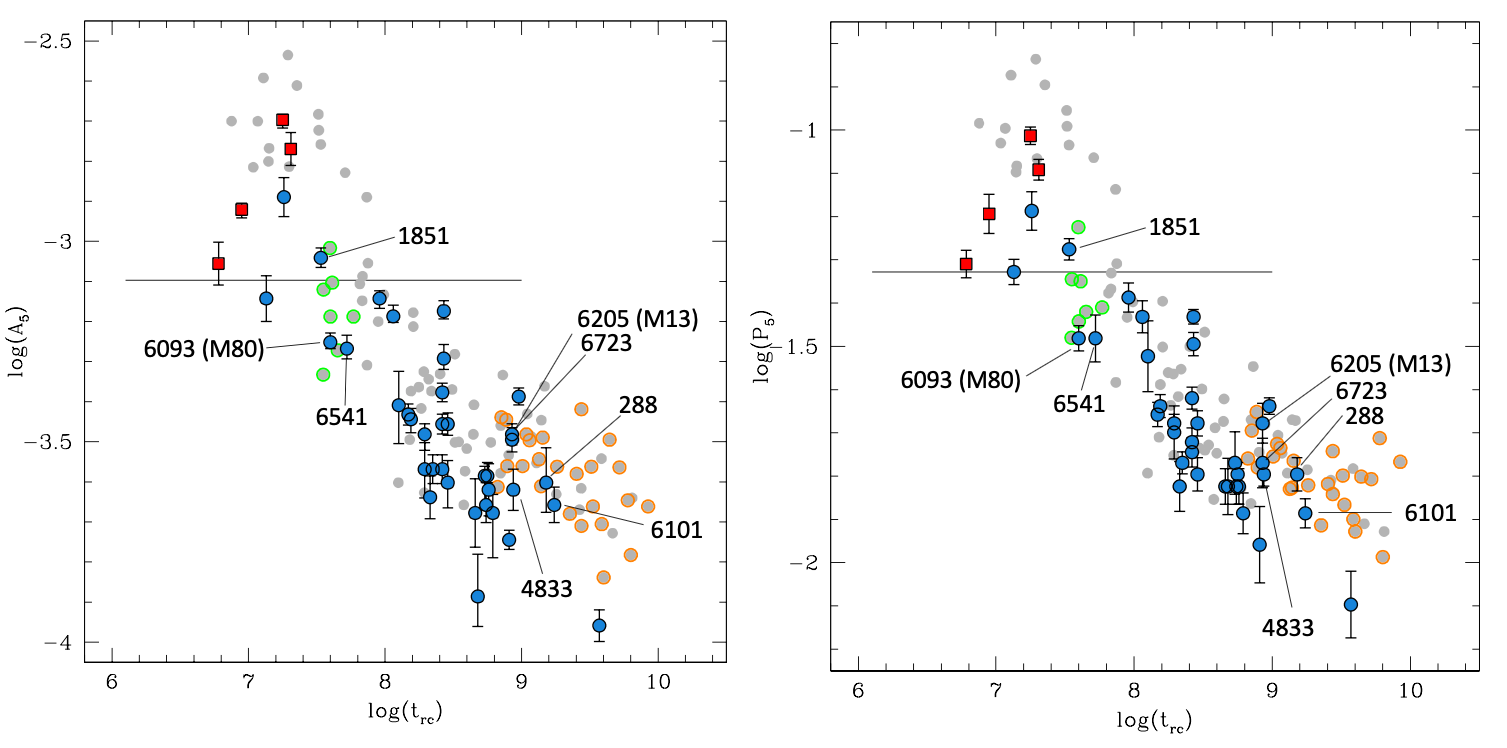}
    \caption{$A_5$ and $P_5$ parameters (right and left panels, respectively) as a function of the central relaxation time, in logarithmic units. The meaning of the symbols and horizontal line is as in Fig. \ref{conc}. The gray symbols encircled in green and orange correspond, respectively, to the simulated clusters that, at an age of 13 Gyr, host a central IMBH (green) and at least 30 stellar-mass BHs (orange).}
\label{trc}
\end{figure*}

Figure \ref{a5p5} shows $P_5$ as a function of $A_5$. The observed values (red and blue symbols) are very well superposed to those measured in the simulated clusters (gray circles), closely following the same monotonically increasing trend. As in the simulations, most of the observed GCs are located in the bottom-left corner of the plot, corresponding to poorly dynamically evolved systems, while the PCC clusters (red squares) are found at systematically larger values of $A_5$ and $P_5$.\footnote{We verified that the same holds if the values of $r_h$ listed in the Holger Baumgardt's online catalog are adopted (see \small{https://people.smp.uq.edu.au/HolgerBaumgardt/globular/parameter.html}).}
The largest values measured in the simulations ($A_5>0.002$ and $P_5>0.1$) are never attained by the observed clusters. As already anticipated by the shape of the nCRDs, NGC 6652 and NGC 1851 have values of $A_5$ and $P_5$ that are very close to those of the PCC systems NGC 6681 and NGC 6624, respectively. The dashed lines in Fig. \ref{a5p5} mark the boundaries of the region that B24 proposed as separation between pre- and post-CC systems. Based on the observational results, we now revise these boundaries, suggesting that the clusters that already experienced the CC phase are characterized by $A_5>0.008$ and $P_5>0.47$. Interestingly, two GCs that are suspected to be PCC systems (namely, NGC 362 and NGC 6717; see \citealp{dalessandro+13} and \citealp{ortolani+99}, respectively) are located very close to these new boundaries. 

In Figure \ref{conc} we study the dependence of $A_5$ on the concentration parameter ($c$) of the King models that best-fit the projected density profiles. As expected, $A_5$ systematically increases with $c$, testifying a higher central segregation of stars in more dynamically evolved systems. However, we notice a disagreement between the observations (colored symbols) and the simulations (gray circles) for $c>1.8$: the observed values of $A_5$ are systematically smaller than those measured in the simulated clusters for any fixed value of $c$, or the observed King concentration is systematically larger than that found in the simulations for fixed $A_5$ (with the only exception of NGC 6652). In addition, while the values of $A_5$ measured in the simulated clusters (gray circles) are systematically larger than those obtained from the direct integration of the 
King models with the same concentration $c$ (solid line), this is no longer the case for the observational data. This once more demonstrates the difficulty and the limits of describing dynamically evolved GCs by using King models, and it indicates that the dependence of $A_5$ on $c$ cannot be efficiently used to pinpoint PCC systems.

Figure \ref{trc} shows $A_5$ and $P_5$ as a function of the central relaxation time ($t_{rc}$) computed by using equation (7) of \citet{djorg93} and listed in Table \ref{tab1}. The well-defined anti-correlations between these parameters once more confirms that $A_5$ and $P_5$ are powerful indicators of dynamical evolution.

Finally, we compare the values of $A_5$ and $P_5$ with those of the $A^+_{rh}$ parameter, which measures the segregation of BSSs toward the center of each system with respect to a reference population of lighter stars (see \citealp{alessandrini+16}). To measure this parameter for Galactic GCs, \citet{ferraro+18} analyzed the same observational data used in this paper, acquired through the HST UV Survey of Galactic GCs \citep{piotto+15}. However, the measure of $A^+_{rh}$ requires a radial sampling of each system out to $r_h$ (while the radial cut adopted to determine $A_5$ and $P_5$ is half this value). Hence, seven GCs in the sample studied here have no value of $A^+_{rh}$ in \citet{ferraro+18}, while the remaining 33 are in common between the two works. Figure \ref{apiu} shows the logarithm of $A_5$ and $P_5$ as a function of $A^+_{rh}$ (left and right panels, respectively). In both diagrams, we find a strong positive correlation, with the most dynamically evolved GCs having the largest values of $A_5$, $P_5$, and $A^+_{rh}$. Interestingly, NGC 362 is close to the boundaries between pre-CC and PCC systems adopted in this work (solid line) and, independently, in \citet[][dotted line, corresponding to $A^+_{rh}=0.3$]{ferraro+23b}, which likely suggests once more that it is close to CC.

\begin{figure*}
    \centering  \includegraphics[width=1\linewidth]{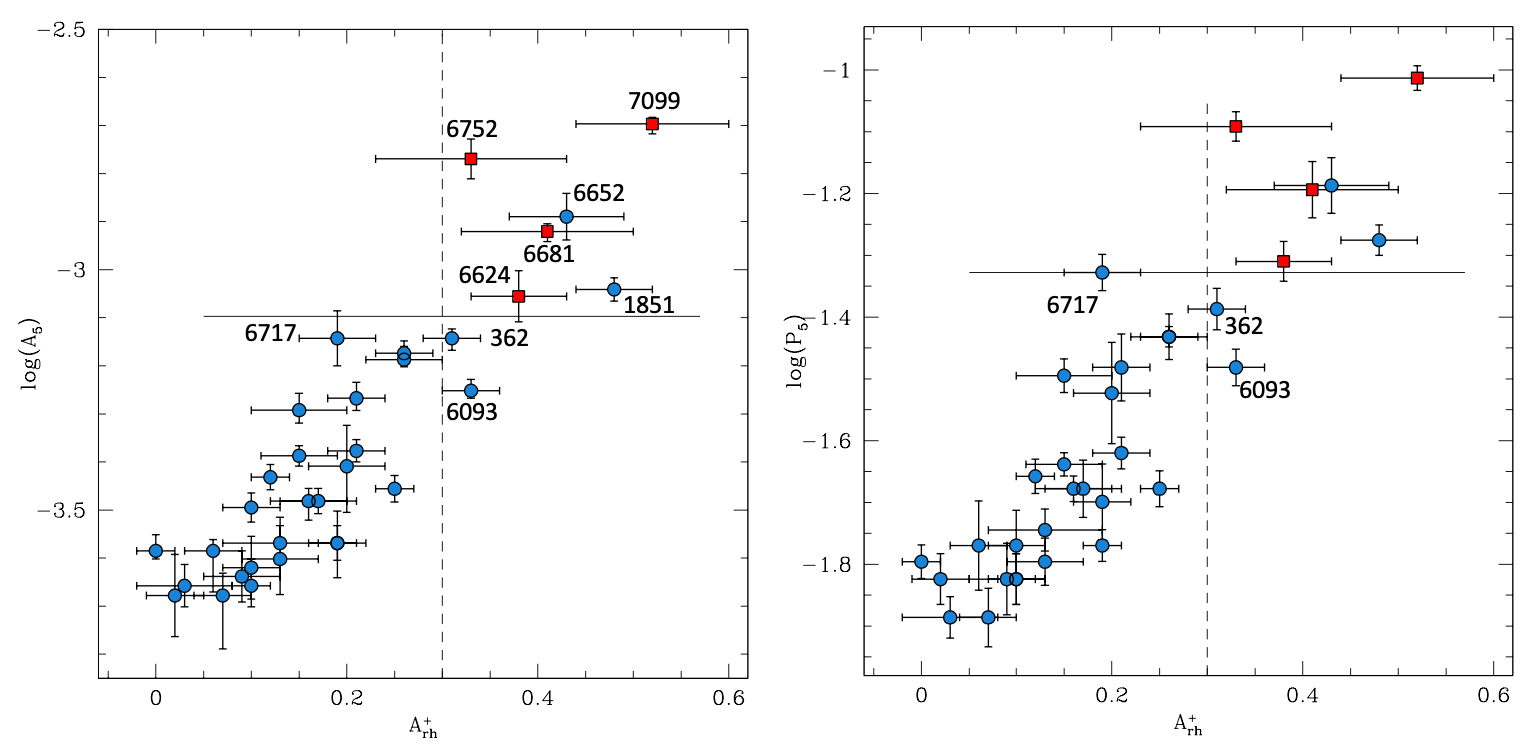}
    \caption{Logarithm of $A_5$ and $P_5$ as a function of the $A^+_{rh}$ parameter (left and right panels, respectively). The solid and dashed lines mark, respectively, the boundaries between pre-CC and PCC systems as proposed in the present work ($A_5>0.008$, $P_5>0.47$), and in \citet[][$A^+_{rh}=0.3$]{ferraro+23b}.}
\label{apiu}
\end{figure*}

\section{Summary and conclusions}
\label{conclu}
In a series of previous papers \citep{bhat+22, bhat+23, bhat+24}, we used a suite of dedicated Monte Carlo simulations to define the nCRD parameters $A_5$ and $P_5$ (see Sect. \ref{para}) and demonstrate their ability to trace the level of internal dynamical evolution experienced by GCs. In this work we have determined $A_5$ and $P_5$ from observational data acquired with HST photometry \citep{piotto+15} for a sample of 40 Galactic GCs. 

Figures \ref{a5p5} and \ref{trc} demonstrate an excellent agreement between observations and simulations. In addition, both parameters show a well-defined anti-correlation with the central relaxation time. These results strongly confirm that $A_5$ and $P_5$ are powerful diagnostics of dynamical evolution. From the observed position of the PCC clusters in the $A_5-P_5$ diagram (red squares in Fig. \ref{a5p5}), we revise the location of the boundaries separating pre-CC from PCC systems with respect to those originally proposed by B24 (dashed lines in Fig. \ref{a5p5}): the comparison with observational data suggests $A_5>0.008$ and $P_5>0.47$ (solid lines in the same figure). However, these should be considered just as reference values since CC is not expected to be an instantaneous phenomenon, and a smooth transition between the pre- and the post-CC states is therefore expected, with a number of clusters being on the verge of starting CC, or currently experiencing it. In fact, clusters like NGC 6717 and NGC 362, which are suspected to be close to reaching CC \citep{ortolani+99, dalessandro+13}, are just below these limits and may indicate the existence of a ``gray region'' of transition between the two states. The comparison shown in Fig. \ref{conc} indicates that higher central concentration systems have larger values of $A_5$, in agreement with the expectations from an effect of internal dynamical evolution. 

The remarkable agreement between observations and simulations in the $A_5-t_{rc}$ diagram (Fig. \ref{trc}) allows us to tentatively identify the GCs that more likely host an intermediate-mass black hole (IMBH) or a large number of stellar-mass BHs. In fact, from the Monte Carlo simulation analysis, B24 noticed that in the $A_5 - t_{rc}$ diagram, the systems including a central IMBH are all clumped at $-3.35 < \log(A_5) < -3.0$ and $\log(t_{rc}) \sim 7.6$ (see the gray symbols encircled in green in Fig. \ref{trc}), and the clusters hosting more than 30 stellar-mass BHs tend to populate the bottom-right corner of the distribution (see the gray symbols encircled in orange). 
As labeled in Fig. \ref{trc}, a few of the Galactic GCs studied in this work occupy the locus of the $A_5-t_{rc}$ and  $P_5-t_{rc}$ planes populated by models containing either an IMBH or more than 30 stellar-mass BHs. Although these portions of the $A_5-t_{rc}$ and  $P_5-t_{rc}$ planes are also populated by models with no IMBH and no stellar-mass BHs, the identification of GCs with these values of $A_5$, $P_5$, and $t_{rc}$ may guide further investigations needed to confirm or rule out the presence of the suspected BHs (based, for example, on the the study of GC internal kinematics; see, e.g., \citealp{dellacroce+24}). Thus in principle, this diagram  provides a useful tool to select potentially interesting clusters, with specific investigations then needed to confirm or exclude the presence of the suspected BHs.

Fig. \ref{apiu} shows that $A_5$ and $P_5$ strongly correlate also with the $A^+_{rh}$ parameter, which traces GC dynamical evolution through the level of central segregation of their BSS populations \citep{ferraro+18}. This further demonstrate the potential of these parameters as diagnostics of internal dynamics. Since each approach is affected by operational difficulties and uncertainties (like any other method proposed in the literature; e.g., \citealt{baum+03, bianchini+16, bianchini+18, tiongco+16, webb+17}), the possibility of using more than one parameter is of great advantage. For instance, $A_5$ and $P_5$ could be particularly useful in clusters with very small populations of BSSs, or where the CMD region of BSSs is strongly contaminated by the host galaxy field stars. The main limitation of the nCRD parameter approach is the need for high-resolution, homogeneous, and complete photometry down to below the MS-TO, from the very center out to $0.5 \times r_h$, in both extended and very concentrated GCs. 
In addition, determining $r_h$ in PCC clusters, where the density profile is not well reproduced by flat-core King models, is not straightforward. Hence, the method would also benefit from a redefinition of the radial cut used to build the nCRDs ($R_{\rm cut}$), avoiding the need of using King model fitting.

\begin{acknowledgements}
This work is part of the project {\it "GENESIS - Searching for the primordial structures of the Universe in the heart of the Galaxy"} (Advanced Grant FIS-2024-02056, PI:Ferraro), funded by the Italian MUR through the {\it Fondo Italiano per la Scienza (FIS)} call.  
\end{acknowledgements}

\centering
\begin{table*}
 \caption{King concentration, half-mass radius, central relaxation time, and nCRD parameters with their uncertainties, for the 40 Galactic GCs studied in this work.}
    \renewcommand{\arraystretch}{1.8}
\begin{tabular}{|l|c|r|c|c|c||l|c|r|c|c|c|}
    \hline
    Name &  $c$ & $r_h~~~~~$ &  $\log(t_{rc})$ & $A_5 \times 10^4$ & $P_5 \times 10^2$ & Name &  $c$ & $r_h~~~~~$ &  $\log(t_{rc})$ & $A_5 \times 10^4$ & $P_5 \times 10^2$ \\
\hline
NGC  288 & 0.98 & 167.20 & 9.18 & $  2.5 _{-0.4 }^{+0.5 }$ & $  1.6 _{-0.1 }^{+0.1 }$  &  NGC 6254 & 1.41 & 139.80 & 8.29 & $  3.3 _{-0.3 }^{+0.2 }$ & $  2.1 _{-0.1 }^{+0.1 }$  \\
\hline										      	  
NGC  362 & 1.73 &  73.84 & 7.96 & $  7.2 _{-0.4 }^{+0.3 }$ & $  4.1 _{-0.3 }^{+0.3 }$  &  NGC 6341 & 1.74 &  85.00 & 8.06 & $  6.5 _{-0.2 }^{+0.4 }$ & $  3.7 _{-0.3 }^{+0.3 }$  \\
\hline										      	  
NGC 1261 & 1.16 &  40.80 & 8.74 & $  2.2 _{-0.2 }^{+0.3 }$ & $  1.5 _{-0.1 }^{+0.1 }$  &  NGC 6352 & 1.10 & 123.00 & 8.46 & $  2.5 _{-0.4 }^{+0.3 }$ & $  1.6 _{-0.1 }^{+0.1 }$  \\
\hline										      	  
NGC 1851 & 1.95 &  51.00 & 7.53 & $  9.1 _{-0.5 }^{+0.5 }$ & $  5.3 _{-0.3 }^{+0.3 }$  &  NGC 6362 & 1.09 & 123.00 & 8.91 & $  1.8 _{-0.1 }^{+0.1 }$ & $  1.1 _{-0.2 }^{+0.2 }$  \\
\hline										      	  
NGC 2298 & 1.38 &  58.80 & 8.10 & $  3.9 _{-0.9 }^{+0.8 }$ & $  3.0 _{-0.6 }^{+0.6 }$  &  NGC 6496 & 1.18 &  93.60 & 8.76 & $  2.4 _{-0.4 }^{+0.4 }$ & $  1.5 _{-0.1 }^{+0.1 }$  \\
\hline										      	  
NGC 2808 & 1.56 &  48.00 & 8.35 & $  2.7 _{-0.2 }^{+0.2 }$ & $  1.7 _{-0.1 }^{+0.1 }$  &  NGC 6541 & 1.86 &  63.60 & 7.72 & $  5.4 _{-0.3 }^{+0.4 }$ & $  3.3 _{-0.4 }^{+0.4 }$  \\
\hline										      	  
NGC 4590 & 1.41 &  90.60 & 8.66 & $  2.1 _{-0.4 }^{+0.4 }$ & $  1.5 _{-0.1 }^{+0.1 }$  &  NGC 6584 & 1.47 &  43.80 & 8.33 & $  2.3 _{-0.3 }^{+0.3 }$ & $  1.5 _{-0.2 }^{+0.2 }$  \\
\hline										      	  
NGC 4833 & 1.26 & 144.60 & 8.94 & $  2.4 _{-0.3 }^{+0.3 }$ & $  1.6 _{-0.1 }^{+0.1 }$  &  NGC 6624 & 2.00 &  38.50 & 6.78 & $  8.8 _{-1.1 }^{+1.1 }$ & $  4.9 _{-0.4 }^{+0.4 }$  \\
\hline										      	  
NGC 5024 & 1.58 &  98.80 & 8.98 & $  4.1 _{-0.2 }^{+0.2 }$ & $  2.3 _{-0.1 }^{+0.1 }$  &  NGC 6637 & 1.38 &  50.40 & 8.17 & $  3.7 _{-0.2 }^{+0.2 }$ & $  2.2 _{-0.1 }^{+0.1 }$  \\
\hline										      	  
NGC 5272 & 1.89 & 138.60 & 8.43 & $  6.7 _{-0.3 }^{+0.4 }$ & $  3.7 _{-0.1 }^{+0.1 }$  &  NGC 6652 & 1.80 &  28.80 & 7.26 & $ 12.9 _{-1.5 }^{+1.4 }$ & $  6.5 _{-0.7 }^{+0.7 }$  \\
\hline										      	  
NGC 5286 & 1.41 &  43.80 & 8.46 & $  3.5 _{-0.2 }^{+0.2 }$ & $  2.1 _{-0.1 }^{+0.1 }$  &  NGC 6681 & 2.07 &  52.10 & 6.95 & $ 12.0 _{-0.6 }^{+0.4 }$ & $  6.4 _{-0.7 }^{+0.7 }$  \\
\hline										      	  
NGC 5897 & 0.86 & 123.60 & 9.57 & $  1.1 _{-0.1 }^{+0.1 }$ & $  0.8 _{-0.1 }^{+0.1 }$  &  NGC 6717 & 1.71 &  45.02 & 7.13 & $  7.2 _{-0.9 }^{+0.9 }$ & $  4.7 _{-0.3 }^{+0.3 }$  \\
\hline										      	  
NGC 5904 & 1.68 & 124.20 & 8.43 & $  5.1 _{-0.3 }^{+0.4 }$ & $  3.2 _{-0.2 }^{+0.2 }$  &  NGC 6723 & 1.11 &  91.80 & 8.93 & $  3.2 _{-0.2 }^{+0.2 }$ & $  1.7 _{-0.2 }^{+0.2 }$  \\
\hline										      	  
NGC 5986 & 1.23 &  58.80 & 8.75 & $  2.6 _{-0.1 }^{+0.2 }$ & $  1.6 _{-0.1 }^{+0.1 }$  &  NGC 6752 & 2.09 & 194.50 & 7.31 & $ 17.0 _{-1.6 }^{+1.6 }$ & $  8.1 _{-0.4 }^{+0.4 }$  \\
\hline										      	  
NGC 6093 & 1.74 &  40.60 & 7.60 & $  5.6 _{-0.2 }^{+0.3 }$ & $  3.3 _{-0.2 }^{+0.2 }$  &  NGC 6779 & 1.38 &  66.00 & 8.42 & $  2.7 _{-0.2 }^{+0.2 }$ & $  1.8 _{-0.1 }^{+0.1 }$  \\
\hline										      	  
NGC 6101 & 1.30 & 128.10 & 9.24 & $  2.2 _{-0.2 }^{+0.2 }$ & $  1.3 _{-0.1 }^{+0.1 }$  &  NGC 6838 & 1.15 & 142.45 & 8.68 & $  1.3 _{-0.2 }^{+0.3 }$ & $  1.5 _{-0.2 }^{+0.2 }$  \\
\hline										      	  
NGC 6144 & 1.55 &  97.80 & 8.73 & $  2.6 _{-0.5 }^{+0.1 }$ & $  1.7 _{-0.3 }^{+0.3 }$  &  NGC 6934 & 1.53 &  41.40 & 8.29 & $  2.7 _{-0.4 }^{+0.4 }$ & $  2.0 _{-0.3 }^{+0.3 }$  \\
\hline										      	  
NGC 6171 & 1.53 & 103.80 & 8.19 & $  3.6 _{-0.3 }^{+0.2 }$ & $  2.3 _{-0.1 }^{+0.1 }$  &  NGC 6981 & 1.21 &  55.80 & 8.79 & $  2.1 _{-0.5 }^{+0.2 }$ & $  1.3 _{-0.1 }^{+0.1 }$  \\
\hline										      	  
NGC 6205 & 1.32 & 148.50 & 8.93 & $  3.3 _{-0.2 }^{+0.2 }$ & $  2.1 _{-0.2 }^{+0.2 }$  &  NGC 7089 & 1.57 &  66.30 & 8.42 & $  4.2 _{-0.2 }^{+0.2 }$ & $  2.4 _{-0.1 }^{+0.1 }$  \\
\hline										      	  
NGC 6218 & 1.34 & 106.20 & 8.42 & $  3.5 _{-0.2 }^{+0.2 }$ & $  1.9 _{-0.1 }^{+0.1 }$  &  NGC 7099 & 2.08 &  91.70 & 7.25 & $ 20.1 _{-0.9 }^{+0.6 }$ & $  9.7 _{-0.4 }^{+0.4 }$  \\
\hline
\end{tabular}
\\
Notes: The half-mass radius ($r_h$) is in arcseconds, the central relaxation time ($t_{rc}$) in years. The values of $A_5$ and $P_5$ plotted in the paper figures correspond to those listed in this table multiplied by $10^{-4}$ and $10^{-2}$, respectively.
\label{tab1}   
\end{table*}

\end{document}